\documentclass[aps,prl,twocolumn,groupedaddress,showpacs]{revtex4}
\usepackage{graphicx}
\graphicspath{{figure file fold path}}
\usepackage[dvipsnames,usenames]{color}
\usepackage{color}
\usepackage{amsmath}
\usepackage{float}
\usepackage{amssymb}
\usepackage{amsthm}
\usepackage{mathrsfs}

\emergencystretch=\maxdimen
\begin{document}

\title{Mode locking via delayed orthogonal-polarization reinjection in semiconductor VCSELs}
\author{T. Wang$^{1}$, Y. Ma$^{1}$, Z. Li$^{1}$, Y. Li$^{1}$, Z. Tu$^{1}$, Y. Zhang$^{1}$, G. Xu$^{2}$, S. Baland$^{3}$, S. Xiang$^{1}$, Y. Hao$^{4}$}

\affiliation{$^1$State Key Laboratory of Integrated Service Networks, School of Telecommunications Engineering, Xidian University, Xi’an 710071, China}
\affiliation{$^2$School of Optical and Electronic Information \& Wuhan National Laboratory for Optoelectronics, Huazhong University of Science and Technology, 1037 Luoyu Road, Wuhan 430074, China}
\affiliation{$^3$Universit\'e C\^ ote d’Azur, Institute de Physique de Nice, UMR 7010 CNRS, Nice, 06200, France}
\affiliation{$^4$State Key Discipline Laboratory of Wide Bandgap Semiconductor Technology, School of Microelectronics, Xidian University, Xi’an 710071, China}

\date{\today}

\begin{abstract}
We demonstrate harmonic mode-locking in a semiconductor VCSEL using polarization-controlled delayed feedback. By integrating a rotatable $\lambda$/2-plate within an external cavity, we achieve precise control over pulse multiplicity and repetition rates in TE and TM modes. For the TE mode, increasing the $\lambda$/2-plate angle ($\theta$) transitions the system from disordered quasi-periodic states to stable fundamental (single-pulse) and harmonic dual-pulse mode-locking. Polarization-resolved measurements and cross-correlation analyses reveal coherent pulse alignment at half the cavity roundtrip time, enabled by polarization-mediated nonlinear dynamics. This work establishes cross-polarization feedback as a fundamental mechanism for ultrafast pulse engineering, advancing the understanding of polarization-mediated nonlinear dynamics in laser physics.  
\end{abstract}


\maketitle 

\section{Introduction}
In recent years, mode locking has been considered as a cornerstone technique for generating ultrashort optical pulses with repetition rates ranging from MHz to GHz~\cite{Gordon2002, keller2021ultrafast}. This capability provides promising platforms for fundamental research and various applications, including optical telecommunications, spectroscopy, material processing and neuromorphic computing.~\cite{Avrutin2000, Julien2006, Aadhi2024, Weid2024}. Consequently, the investigation of mode locking in semiconductor lasers has garnered significant attention from researchers.

Mode locking is a resonant process that generates pulsed radiation by synchronizing the phases of multiple longitudinal modes within the resonator~\cite{Zhan2007}. Typically, it can be achieved actively~\cite{Revin2016} or passively~\cite{Marconi2014}. In this work, we focus on passive mode-locking, which doesn't need for any external modulation, making it the preferred method for generating optical pulses at multi-GHz repetition rates~\cite{Yadav2023}. This regime is typically achieved by integrating two key components: a laser amplifier that supplies gain and a saturable absorber that functions as a pulse-shortening element~\cite{Marconi2015}. 

Apart from inserting a saturable absorber~\cite{ruschel2025regenerative}, crossed-polarization gain modulation method was proposed to achieve passive mode-locking for semiconductor lasers~\cite{Julien2006}, However, experimental validation of this concept in its most straightforward implementation remains elusive. Here, inspired by the theoretical prediction~\cite{Julien2006}, we introduce a novel approach to harmonic mode-locking in VCSELs using delayed orthogonal-polarization reinjection. By rotating a $\lambda/2$-plate within an external feedback cavity, we dynamically control the interplay between TE and TM modes, enabling synchronized pulse generation in both polarizations. Unlike conventional methods, our scheme eliminates the need for saturable absorbers or active modulation, leveraging polarization-state matching to regulate pulse interactions. Thus, our work not only confirms the feasibility of polarization-mediated mode-locking but also reveals a novel subharmonic synchronization mechanism. These findings redefine polarization as a dynamic control parameter in ultrafast laser physics, offering a versatile platform for tailoring nonlinear photonic systems.

\section{Experimental setup}
\begin{figure}[ht!]
\centering
  \includegraphics[width=7.5cm]{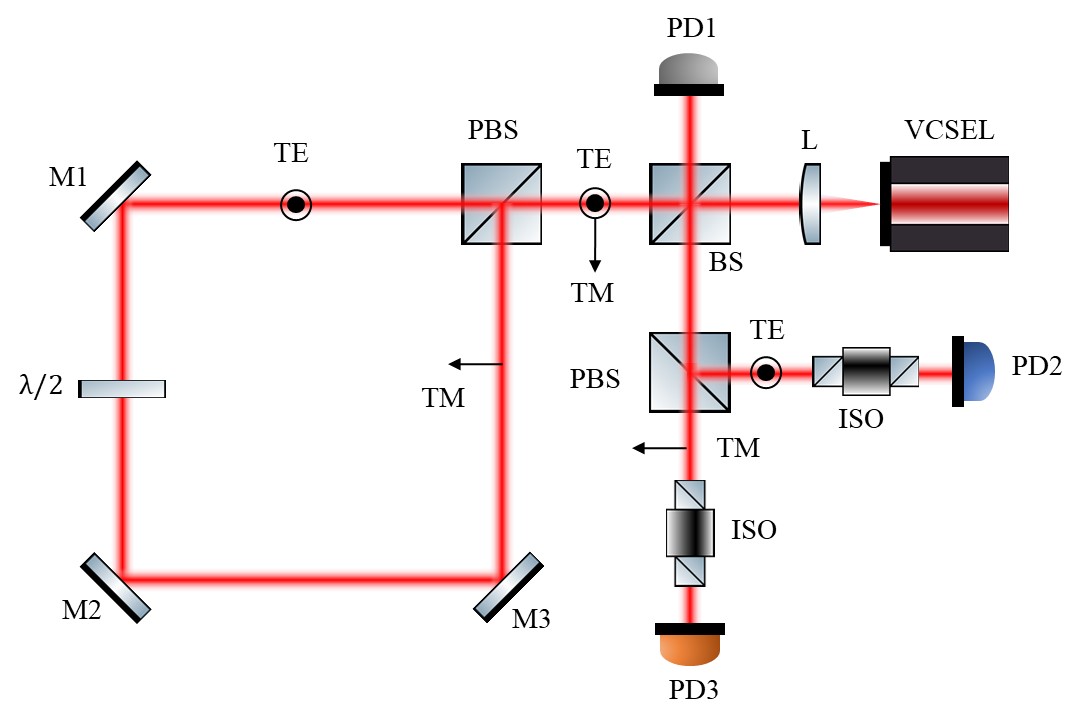}
  \caption{Experimental setup: VCSEL, Vertical Cavity Surface Emitting Laser; L, optical lens; BS, beam splitter; PBS, polarized beam splitter; M, mirror; $\lambda/2$, half-wave plate; ISO, optical isolator; PDn, photodetectors; Osc., oscilloscope.
}
  \label{Setup}
\end{figure}

The experimental configuration (Fig.~\ref{Setup}) employs a semiconductor VCSEL (Thorlabs L850VH1) with emission wavelength $\lambda = 850$ nm, and which is stabilized at 25$^\circ$C by a temperature controller (Thorlabs TED200C). The VCSEL operates in a single longitudinal and transverse mode, as well as in one of two orthogonal linear polarization modes, and is driven by a low-noise current source (Thorlabs LDC205C). No
$\lambda/2$-plate is placed between the VCSEL and the non-polarizing beam splitter (BS), as the polarization-resolved L-I curves (Fig.~\ref{S-curve-average-power}a) confirm that the TE/TM modes’ intrinsic principal axes align naturally with the PBS axes. This deliberate alignment - achieved through pre-experimental calibration, the setup’s integrity and the results’ interpretability.

After passing through the non-polarizing BS, the laser light is sent to an external ring cavity, which can reflect the light back. A polarizing beam splitter (PBS) and a $\lambda/2$-plate are placed within the external cavity. Inside the cavity, the PBS separates the TE and TM modes, which then can undergo up to 90-degree rotation after passing through the $\lambda/2$-plate. The second arm exiting the first BS is directed towards another PBS for detection. Both TE and TM outputs of the PBS are directed towards optical isolators and finally two AC-coupled and 10 GHz - bandwidth amplified fast detectors (PD2 and PD3). The temporal signals are then monitored using a 20 GHz oscilloscope (Tektronix DSA72004) with a sampling rate of 50 GHz. The total feedback length within the system is 1.8 meters, corresponding to a round-trip time of approximately 12 ns.

\section{Results and discussions}
\begin{figure}[ht!]
\centering
  \includegraphics[width=7.5cm]{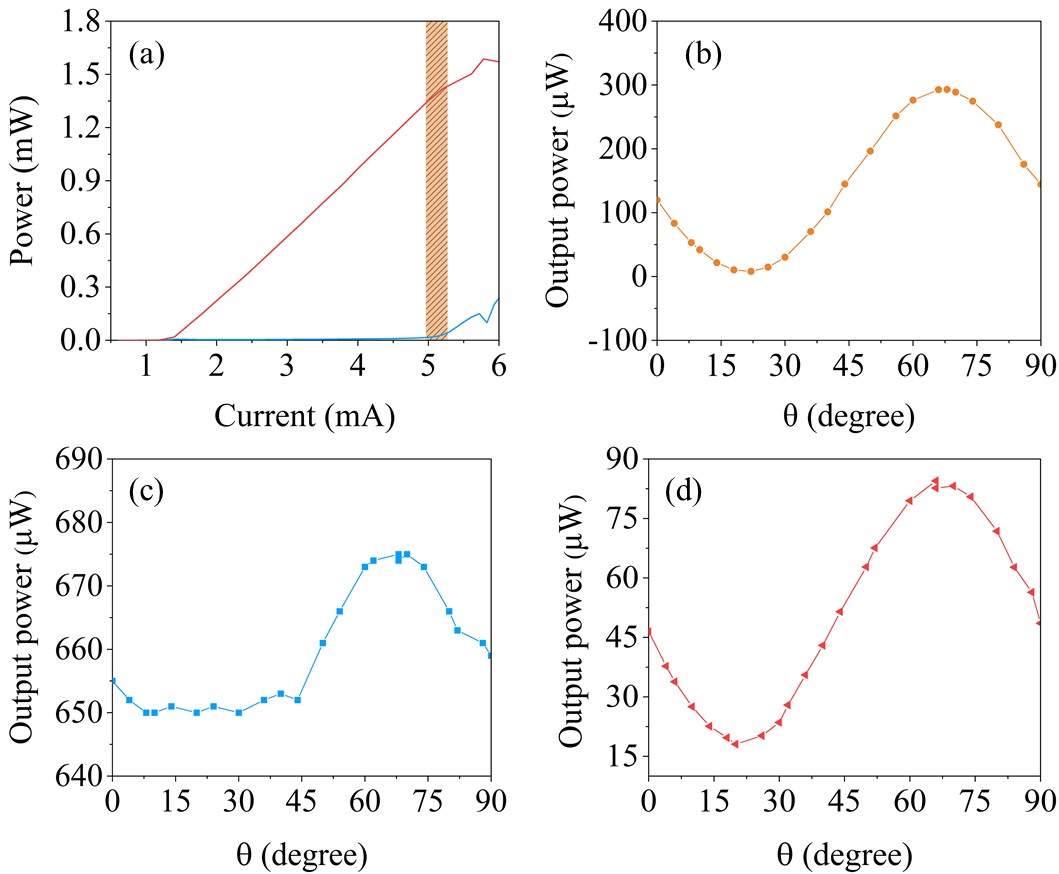}
	\caption{Fundamental characterizations of the the VCSEL under free running and with external feedback: (a) input-output function curves of the TE and TM modes as a function of pump current. The light blue curve indicates the TE mode, and the red curve denotes the TM mode; (b) averaged power of the feedback light, measured at 5.30 mA, after circulating from the ring cavity; (c) and (d), output powers of the TE and TM modes at $J = 5.30$ mA after the injection of the feedback light.}
  \label{S-curve-average-power}
\end{figure}

The free-running VCSEL exhibits distinct thresholds for TE ($J_{th}^{TE} \approx 1.48 mA$ ) and TM ($J_{th}^{TM} \approx 5.20 mA$) modes (Fig.~\ref{S-curve-average-power}a).
In the intermediate current regime ($J_{th}^{TE} < J < J_{th}^{TM}$), the output of the TE mode linearly increases with injection current (red curve), while the power of the TM mode remains almost zero (light blue curve). The TM mode starts to be lasing when $J > J_{th}^{TM}$, and polarization dynamics is observed for both modes. The orange-shaded region near $J_{th}^{TM}$ holds particular significance, as enhanced orthogonal-polarization gain modulation in this regime – consistent with theoretical frameworks ~\cite{Julien2006, masoller1999} -- creates favorable conditions for passive mode-locking through nonlinear polarization coupling~\cite{Fermann1993}. 

Fig.~\ref{S-curve-average-power}b shows the variation in the feedback light power (measured by the power meter PD1) as a function of $\theta$ at $J = 5.30$ mA. We underline that $\theta$ refers to the wave plate orientation with respect to an arbitrary origin and empirically we observe that the reinjection is minimum near $\theta=24^\circ$ and maximal near $\theta=70^\circ$ (see Fig.\ref{S-curve-average-power}b). More information can be found in the polarization-resolved time-averaged intensities, which show that the intracavity nonlinear interaction induces distinct mode-selective responses: (i) the TE-mode output (measured by a power meter in the location of PD2) remains constant at 650 $\mu$W for $\theta < 45^\circ$ which indicates that up to that angle, the reinjection has very moderate effect on the time-averaged TE intensity. At higher angles the time-averaged TE intensity grows by up to 5$\%$ increase, peaking at $\theta < 65^\circ$ (Fig.~\ref{S-curve-average-power}c), suggesting constructive interference in dominant polarization channel; (ii) the TM-mode output (measured at the location of PD3) conversely exhibits in-phase modulation with feedback signal but at much lower intensity (Fig.~\ref{S-curve-average-power}d), indicative of weak cross-polarization coupling. These observations highlight the critical role of polarization-state matching in governing nonlinear mode competition dynamics.

\begin{figure*}[ht!]
\centering
  \includegraphics[width=13.5cm]{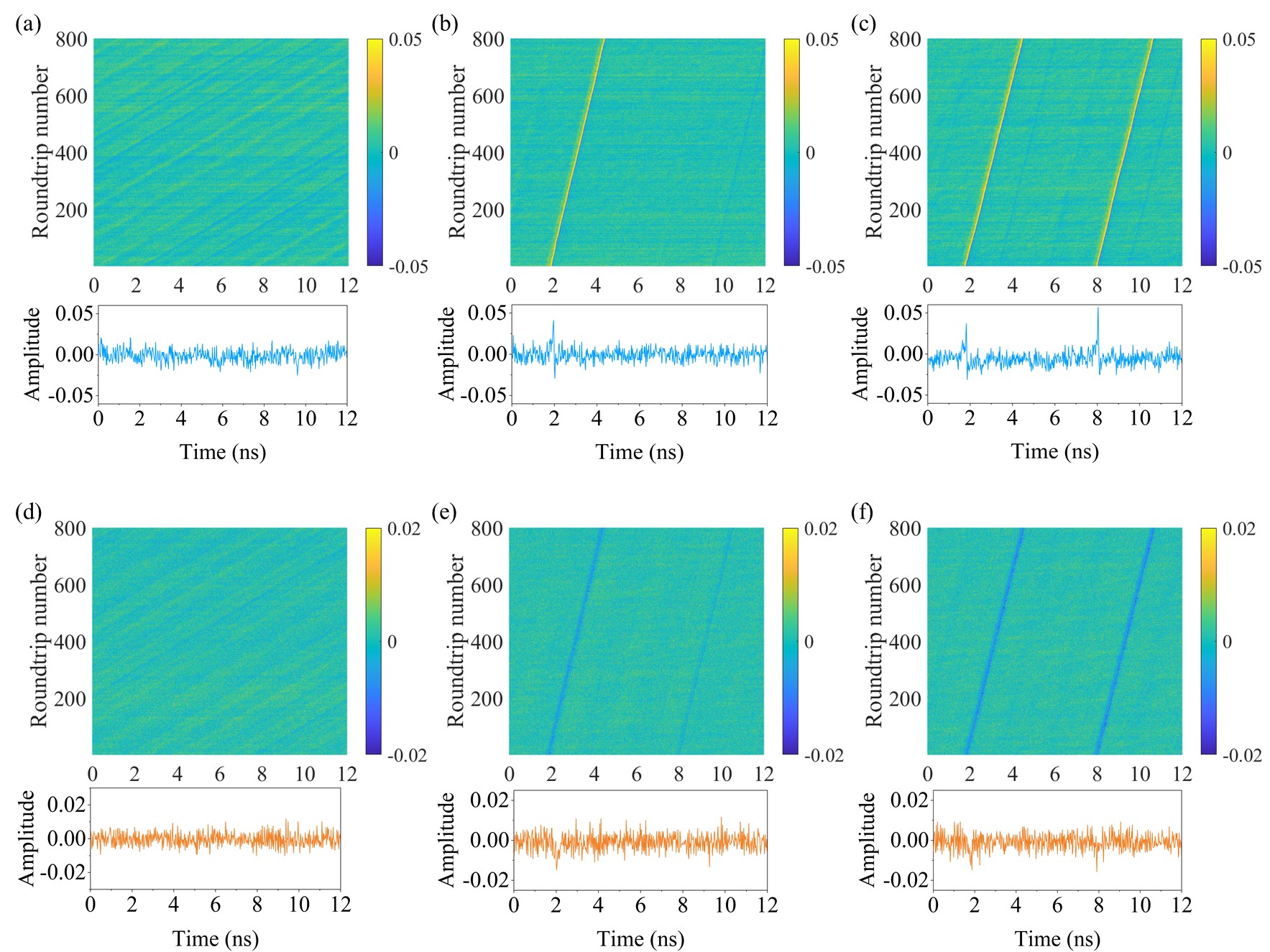}
  \caption{Spatio-temporal diagrams and temporal dynamics of the TE and TM modes in the laser output under a bias current $J = 5.30$ mA, with $\lambda$/2-plate angles $\theta = 50^\circ$ (a and d), $60^\circ$ (b and e) and $70^\circ$ (c and f), respectively. a-c for the TE mode; d-f for the TM mode.}
  \label{Temporal-spatial-merge}
\end{figure*}

Fig.~\ref{Temporal-spatial-merge}a-c presents the spatio-temporal diagrams and typical dynamics of the TE mode under a bias current of $J = 5.30$ mA with $\theta$ values of  $50^\circ$, $60^\circ$ and $70^\circ$, respectively. It is clear that the $\lambda$/2-plate plays significant role on the mode dynamics. At $\theta = 50^\circ$ (Fig.~\ref{Temporal-spatial-merge}a), a weakly modulated periodic pattern appears, characterized by small-amplitude oscillations within each roundtrip ($T_{rt}$), as shown by the bottom panel of Fig.~\ref{Temporal-spatial-merge}a. Increasing $\theta$ to $60^\circ$ (Fig.~\ref{Temporal-spatial-merge}b) triggers a fundamental mode-locking state, where a single pulse is stably confined per roundtrip (bottom panel of Fig.~\ref{Temporal-spatial-merge}b). The single-pulse locking may come from balanced nonlinear phase shifts and gain-loss equilibrium, enabled by the plate’s orientation~\cite{Grelu2012, Zhao2009}.

Further increasing $\theta = 70^\circ$ (Fig.~\ref{Temporal-spatial-merge}c) results in dual-pulse locking within one roundtrip $T_{rt}$. As stated above, the origin of $\theta$ is arbitrary and $\theta=70^\circ$ corresponds to the maximal reinjection of TE into TM as visible on Fig.~\ref{S-curve-average-power}b (situation which can also be achieved at $\theta=-20^\circ$). The asymmetry of the pulses as measured by the AC-coupled detectors suggest slightly different rising and falling slopes. We underline that very precise tuning of the half-wave plate is required to reach this regime, suggesting the key role of precise feedback phase. As we discuss later on, this is a case of harmonic mode-locking since both pulses are evenly distributed within one roundtrip, so that the period of this solution is exactly half that of the fundamental mode-locked stated. 


Fig.~\ref{Temporal-spatial-merge}d-f exhibits the corresponding spatio-temporal diagrams and representative dynamics of the TM mode under the same bias current and the same angles of the $\lambda$/2-plate with the TE mode, respectively. At $\theta = 50^\circ$ (Fig.~\ref{Temporal-spatial-merge}d), the periodic structure is less pronounced compared to the TE mode, though visible temporal features emerge within a single roundtrip (bottom panel). Increasing $\theta$ to $60^\circ$ (Fig.~\ref{Temporal-spatial-merge}e) stabilizes a harmonic mode-locking state, where two pulses (left pulse stronger than the right) are confined per roundtrip - a notable contrast to the single-pulse locking observed in the TE mode (bottom panel of Fig.~\ref{Temporal-spatial-merge}e). Besides, we also observe that both pulses exhibit significantly weaker peak amplitudes than their TE counterparts (lower panel).

Further increasing $\theta$ to $70^\circ$ leads to stable dual-pulse locking, but with comparable amplitudes (Fig.~\ref{Temporal-spatial-merge}f), mirroring the TE mode’s behavior at this angle. Obviously, the amplitudes of both pulses are much weaker on the bottom panel of Fig.~\ref{Temporal-spatial-merge}f, highlighting mode-specific nonlinear absorption.



\begin{figure}[ht!]
\centering
  \includegraphics[width=7.3cm]{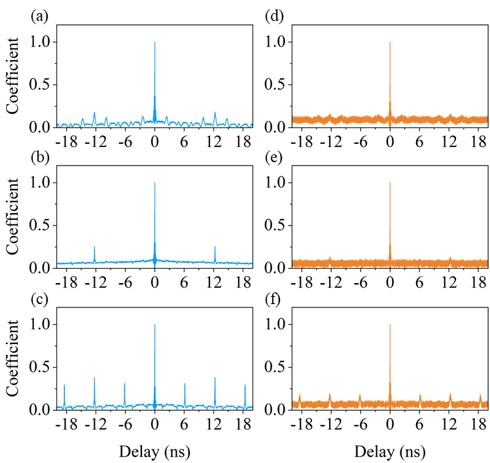}
  \caption{Autocorrelation coefficients calculated by using the temporal signals of the TE and TM mode with different $\lambda$/2-plate angles: (a)-(c) for TE mode; (d)-(f) for TM mode. (a) and (d) for $50^\circ$; (b) and (e) for $60^\circ$; (c) and (f) for $70^\circ$. }
  \label{Auto}
\end{figure}

Further insight can be gained by comparing autocorrelation of each polarization and cross-correlation between both, as shown on Fig.~\ref{Auto}. Specifically, Fig.~\ref{Auto}a-c show the autocorrelation traces of the TE mode with $\lambda$/2-plate angle $\theta = 50^\circ$, $60^\circ$, and $70^\circ$. At $\theta = 50^\circ$ (Fig.~\ref{Auto}a), the trace features a prominent central peak at zero delay with a few irregular sidebands up to 12.4~ns. These features are consistent with the quasi-periodic and weakly modulated patterns in the spatio-temporal representation in Fig.~\ref{Temporal-spatial-merge}a. Increasing $\theta$ to $60^\circ$ (Fig.~\ref{Auto}b), leads to disappearance of the sidebands, leaving only a single secondary peak at 12.3 ns (roundtrip time) which shows fundamental mode-locking with one pulse per roundtrip. At $\theta = 70^\circ$ (Fig~\ref{Temporal-spatial-merge}c), the autocorrelation shows the emergence of a secondary peak at half the round-trip time. The absence of any other feature in the autocorrelation confirms the harmonic locking situation, also indicates the critical role of polarization alignment and feedback phase control in achieving stable mode-locking.

Fig.~\ref{Auto}d-f show the TM mode’s autocorrelation traces under identical $\theta$ values. The TM mode exhibits weaker amplitudes ($\sim 30\%$) and an elevated noise floor, attributed to polarization-dependent losses and mode competition, in agreement with the time-averaged intensities in Fig.~\ref{S-curve-average-power}. At $\theta = 50^\circ$ (Fig.~\ref{Auto}d), broadened peaks and suppressed sidebands indicate polarization-filtered gain saturation, attenuating nonlinear interactions. For $\theta = 60^\circ$ (Fig.~\ref{Auto}e), two feedback signatures emerge: a primary peak at 12.3 ns and a weaker peaks at 6.1 ns and 18.4 ns, suggesting harmonic mode-locking and emergence of TM-specific cavity instabilities. At $\theta = 70^\circ$ (Fig.~\ref{Auto}f), harmonic mode-locking becomes clear, with periodic peaks spaced by 6.15 ns and uniform amplitudes, reflecting symmetrical dual-pulse circulation stabilized by polarization-optimized phase matching.

\begin{figure}[ht!]
\centering
  \includegraphics[width=8.0cm]{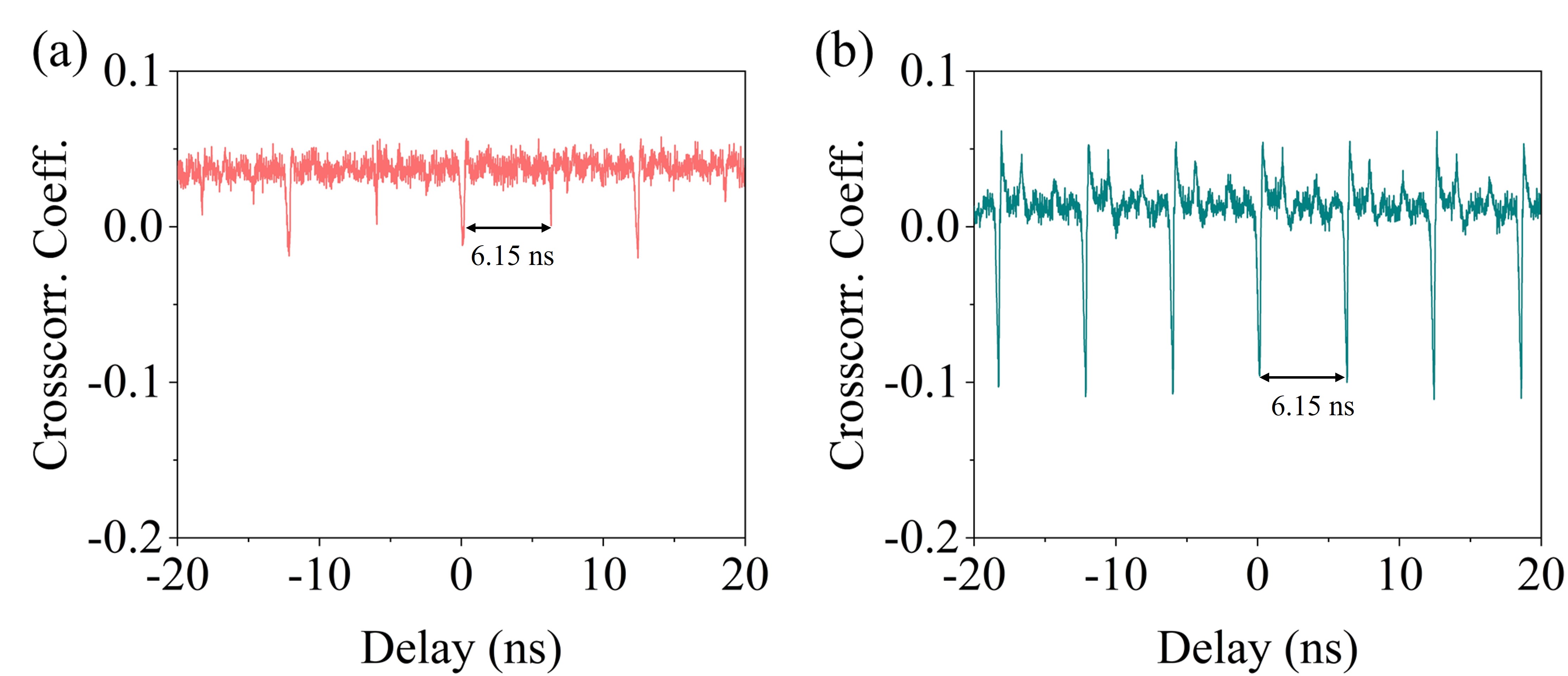}
  \caption{Cross-correlation coefficients calculated by using the temporal signals of the TE and TM mode with different $\lambda$/2-plate angles: (a) $60^\circ$; (b) $70^\circ$. }
  \label{Crosscorrelation} 
\end{figure}

To further understand the polarization-dependent mode-locking dynamics, we calculated the cross-correlation traces between TE and TM modes at $\lambda$/2-plate angle $\theta = 60^\circ$ and $70^\circ$. As shown by Fig.~\ref{Crosscorrelation}a, at $\theta = 60^\circ$, the cross-correlation coefficient exhibits strong anti-correlation (as was also found in chaotic regimes in \cite{nazhan2016chaos}), which originate from polarization-mediated gain competition and anti-phase pulse alignment. These features are superimposed on a pronounced noise floor, reaching 
$\sim 40\%$ of the central peak amplitude, suggesting partially randomized mode coupling. The bimodal amplitude distribution—sharp higher-amplitude peaks (TE-dominated pulses) alternating with suppressed lower-amplitude peaks (TM-dominated)-reflects asymmetric nonlinear coupling dynamics. This asymmetry stems from polarization-dependent gain saturation that selectively amplifies the TE mode, effectively suppressing TM mode oscillations. The observed 
$\sim 6.15$ ns temporal separation between adjacent peaks further suggests periodic energy exchange (anti-phase intensity modulation) between polarization states during pulse evolution.
We underline that the pulse we observe (although through an AC-coupled detector) exist on top of a non-zero background. This suggests that all modes involved do not have the same amplitude (as is the case of the most well-known mode-locking situation). Nevertheless, the sharp pulses involving about 190 modes (computed from mode spacing and a pulse duration of 26 ps) and sharp autocorrelation indicate that all modes are locked to each other with fixed phase relation.

Increasing $\theta$ to $70^\circ$ (Fig.~\ref{Crosscorrelation}b) stabilizes dual-pulse harmonic locking with periodic peaks at $\pm 6.15$  ns (half the cavity round-trip time) and deepened anti-correlation (minimum $\sim -0.1$). This subharmonic regime forms a coherent waveform from two pulses per round-trip, likely driven by resonant interactions between cavity birefringence and TE relaxation oscillations~\cite{Wang2025}. Deviations from this angle induce complex spiking dynamics, underscoring the critical role of empirical optimization in nonlinear polarization control. While cumulative intensity-dependent birefringence may emulate explicit feedback delays akin to Ref.~\cite{Julien2006}, integrating this effect into theoretical frameworks remains essential. The transition from disordered anti-correlation at $60^\circ$ to periodic harmonic locking at $70^\circ$ underscores the critical role of polarization control in achieving coherent multipulse interactions.

\section{Conclusions}
In conclusion, we have presented an experimental demonstration of  fundamental and harmonic mode-locking in semiconductor lasers induced by crossed-polarization gain modulation. Although slightly different from the inspiring work \cite{Julien2006}, our work validates the approach of polarization reinjection for passive mode locking in VCSELs. In our scheme, by rotating the $\lambda$/2-plate, we can precisely control over pulse multiplicity, symmetry, and temporal coherence in both TE and TM modes. The TE mode transitions from weakly modulated quasi-periodic states ($\theta = 50^\circ$) to fundamental mode-locking ($\theta = 60^\circ$) and dual-pulse regimes (
$\theta = 70^\circ$), while the TM mode exhibits subharmonic locking and intermittent dynamics, highlighting polarization-divergent nonlinearities. This approach enables ultrafast sources for photonic neural computing and polarization encoded communications, with harmonic locking (half roundtrip time) offering a route to high-repetition pulses without cavity modification. Our work validates delayed orthogonal-polarization feedback as a novel mechanism for harmonic mode-locking, underscores the critical role of polarization-state control in ultrafast pulse engineering, and advances fundamental insights into laser nonlinear dynamics, paving the way for tailored polarization-driven laser architectures.

\section*{Acknowledgment}
This work is partially supported by National Natural Science Foundation of China (Grant No. 62475206 and 61804036), Key Research and Development Plan of Shaanxi Province of China (Grant No. 2024GH-ZDXM-42), National Key Research and Development Program of China (Grant No. 2021YFB2801900, 2021YFB2801901, 2021YFB2801902, and 2021YFB2801904).



\bibliography{biblio}

\end{document}